\documentstyle[epsf,aps,multicol]{revtex}

\begin{document}
\draft

\title{\Large\bf Simulation Studies on the Stability \\
of the Vortex-Glass Order}

\author{\bf Hikaru Kawamura}

\address{Faculty of Engineering and Design, 
Kyoto Institute of Technology, Kyoto 606-8585, Japan}

\maketitle

\begin{abstract}
The stability of the three-dimensional vortex-glass order in 
random type-II superconductors with point disorder 
is investigated
by equilibrium Monte Carlo simulations based on a lattice 
{\it XY\/} model
with a uniform field threading the system.
It is found that the vortex-glass order, which stably exists 
in the absence of screening, is  destroyed by the screenng 
effect, corroborating the previous finding based on the 
spatially isotropic gauge-glass
model. Estimated critical exponents, however,
deviate considerably  from the values reported for 
the gauge-glass model.
\end{abstract}

\begin{multicols}{2} \narrowtext

Due to the enhanced effect of fluctuations,  
the problem of
the phase diagram of high-$T_c$ superconductors
in applied magnetic fields is highly nontrivial and has attracted
much  interest recently.
For random type-II superconductors with point disorder,
possible existence of an equilibrium thermodynamic phase
called the vortex-glass (VG) phase, where the vortex
was pinned on a long length scale by randomly distributed 
point-pinning centers,
was proposed \cite{Fisher}. In such a VG state,
the phase of the condensate wavefunction is frozen in time 
but  randomly in space, with a
vanishing linear resistivity $\rho _L$.
It is a truly superconducting state separated from
the vortex-liquid phase with a nonzero $\rho _L$ via a continuous
VG transition.

This proposal was  supported by
subsequent experiments.  In particular,  transport measurements on
films \cite{film}
and twinned single 
crystals \cite{crystal} gave  
evidence for the occurrence of a continuous transition into
the glassy superconducting state. 
It should be noticed, however,  that these samples 
often contain extended defects with correlated
disorder, such as grain boundaries, twins and dislocations.
Since these extended defects could pinn the vortex more efficiently
than point defects,
the  possibility still remains that 
extended defects play a crucial 
role in the experimentally observed ``VG transitions'', 
and the sample only with 
point defects  behaves differently [4-7].

Stability of the hypothetical VG state 
was also studied by numerical simualtions. Here,
it is essential to obtain the data of 
{\it appropriate thermodynamic quantities in true equilibrium\/} and to
carefully analyze its size dependence.
So far, such calculations have been limited almost exclusively to 
a highly simplified model called the gauge-glass model.
Previous simulations on the three-dimensional (3D) 
gauge-glass model have indicated 
that, while
the stable VG phase exists in the absence of
screening \cite{Huse,Reger},
the finite screening effect inherent to real superconductors
eventually destabiizes it \cite{Young,Rieger}.

Meanwhile, it has been recognized  that the gauge-glass model
has some obvious drawbacks \cite{Huse}. 
First, it is a spatially isotropic model
without a net field threading the system, in contrast to the reality.
Second, the source of
quenched randomness is artificial. 
The gauge-glass model is a random flux model where the qeunched
randomness appears in the phase factor assoicated with the flux.
In  reality, the quenched component of the flux
is uniform, nothing but the external field, and the quenched
randomness occurs in the superconducting coupling.
It remains unclear whether these simplifications underlying
the gauge-glass model really unaffect the basic physics of the
VG ordering in 3D.

The purpose of the presnet letter is to introduce a model in which
the above  limitations of the gauge-glass model are cured,
and to examine by extensibe Monte Carlo (MC) simulations the
nature of the 3D VG ordering with and without screening
beyond the gauge-glass model.
It is found that, as in the gauge-glass model, the VG phase is stable
in the absence of screening but the finite screening effect
destabilizes it. 

We consider  the  dimensionless Hamiltonian[12,13], 

\begin{eqnarray}
{\cal H}/J = - \sum _{<ij>} J_{ij}\cos (\theta _i-\theta _j-A_{ij}) 
\nonumber \\
+ \frac{\lambda_0^2}{2} 
\sum _p (\vec \nabla \times \vec A -\Phi_{ext})^2, 
\end{eqnarray}
where $J$ is the typical coupling stregth, 
$\theta _i$ is the phase of the condensate
at the $i$-th site of a simple cubic lattice,
$\vec A$ is the fluctuating gauge potential at each link
of the lattice,
the lattice curl $\vec \nabla \times \vec A$
is the directed sum of $A_{ij}$'s around
a plaquette with $A_{ji}=-A_{ij}$, and $\lambda_0$ is the
bare penetration depth in units of lattice spacing.
$\Phi _{ext}$ is an exernal flux threading the elementary 
plaqutte $p$, which is equal to $h$ if the plaquette is on the 
$xy$-plane and zero otherwise, {\it i.e.\/}, a
uniform field is applied along the $z$-direction.
The first sum in (1)
is taken over all nearest-neighbor pairs, while the
second sum over all elementary plaquettes.
Fluctuating  variables to be summed over are the phase variables,
$\theta _i$, at each site and the gauge variables, $A_{ij}$, at each
link. In order to allow for the flux penetration into the 
system, we impose
free boundary conditions 
in all directions 
for both $\theta_i$ and $A_{ij}$ [12,13].
Quenched rondomness occurs only in the superconducting coupling
$J_{ij}$ which is assumed to
be an independent
random variable uniformly distributed between [0,2].
We stress that the aforementioned 
drawbacks of the
gauge-glass model have been cured now:
The present model has  a uniform field
threading the system and the quenched 
randomness occurs in the superconducting coupling, not in the
gauge field.

In addition to the  global U(1) gauge symmetry,
the Hamiltonian (1)  has a {\it local\/} gauge symemtry,
{\it i.e.,\/} the invariance under the local transformation
$\theta_i\rightarrow \theta_i+\Delta$ and 
$A_{i,i+\delta}\rightarrow A_{i,i+\delta}+\Delta$
for an arbirtary  site $i$
($A_{i,i+\delta}$'s are all link variables 
emanating from the site $i$).
We  adopt the Coulomb gauge, 
imposing the condition, ${\rm div} {\bf A}\equiv 
\sum _\delta A_{i,i+\delta}=0$,
at every site $i$. 
In the limit of vanishing screening  $\lambda _0\rightarrow \infty $,
the link variable $A_{ij}$ is quenched to the external-field
value, and 
the fluctuating variable becomes 
the phase variable $\theta _i$ only.

Simulation is performed based on the exchange MC method
where the systems at neighboring temperatures are occasionally
exchanged \cite{huku}. 
We run two independent sequences of  systems 
(replica 1 and 2) in parallel, and
compute a complex overlap $q$ between the local superconducting
order parameters of the two replicas $\psi_i^{(1,2)}\equiv 
\exp (i\theta _i
^{(1,2)})$, 
\begin{equation}
  q = \frac{1}{N}\sum_{i}\psi_{i}^{(1)*}\psi_{i}^{(2)},
\end{equation}
where the summation is taken over all $N=L^3$ sites.
In terms of the overlap $q$, the
Binder ratio 
is calculated by
\begin{equation}
  \label{Binder}
  g(L)=2-\frac{[\langle |q|^4\rangle]}
    {[\langle  |q|^2\rangle]^2},
\end{equation}
where $\langle\cdots\rangle$ represents the thermal average 
and $[\cdots ]$ 
represents the average over bond disorder. 
Note that, thanks to the Coulomb-gauge condition, 
the superconducting order parameter, 
which is
originally not local-gauge invariant, becomes 
a nontrivial quantity.

We deal mainly with two cases;
[I] no screening corresonding to $\lambda_0=\infty $,
and [II] finite screening corresponding to $\lambda_0=2 $,  whereas
some data are taken for the case of stronger screening 
corresponding to $\lambda_0=1$.
In either case, we fix the field intensity to $h=1$ which
corresponds to $f=1/(2\pi )\simeq 0.159 $ flux quanta per plaquette.
We have  chosen the fractional value of $f$ to avoid
the commensurability effect associated with the 
vortex-lattice formation.
The lattice sizes  studied are
$L=6,8,10,12$ and 16 ($\lambda_0=\infty $), and 
$L=6,8,10,12$ ($\lambda_0=2 $).
Equilibration is checked by monitoring the 
stability of the results
against at least three-times longer runs for a subset of samples.
Sample average is taken over   300 ($L=6$), 
200-300 ($L=8$), 120 ($L=10$),
75-150 ($L=12$) and 136 ($L=16$) independent bond realizations.

We begin with the case of no screening 
($\lambda _0=\infty $).
The size and temperature dependence 
of the calcualted Binder ratio
is shown in Fig.1(a).
As can be seen from Fig.1(a),
$g(L)$ for different $L\geq 8$ tends to merge, or weakly cross,
at $T/J=0.68\pm 0.02$, indicating that the VG transition occurs
at a finite temperature in the absence of screening. 
Observed near 
marginal behavior  
suggests that $D=3$ is close to the lower critical dimension.
\begin{figure}[h]
\label{fig:g-noscreening}
\epsfxsize=\columnwidth\epsfbox{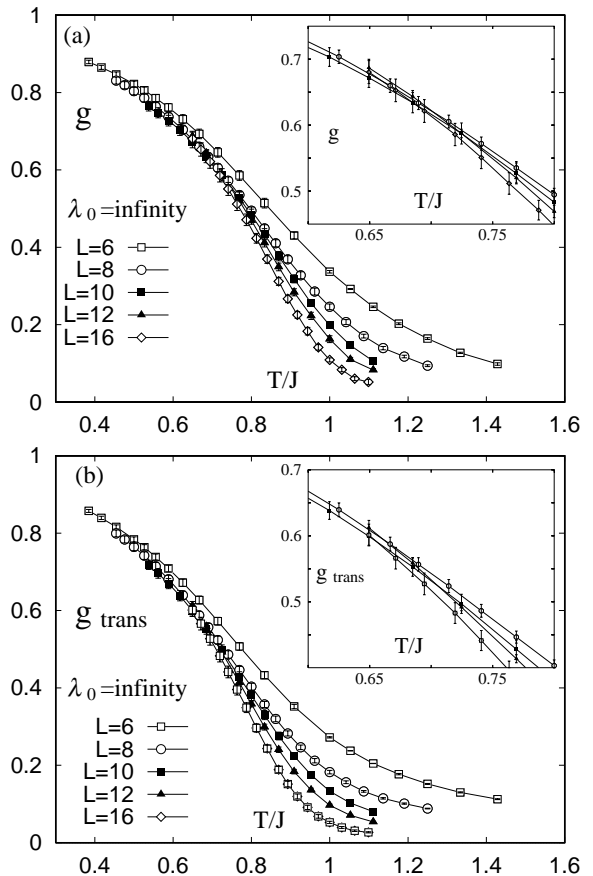}
\caption{
Temperature and size dependence of 
(a) the bulk Binder ratio, Eq.(3),
and of (b)  the transverse Binder ratio, Eq.(5), of 
the model without
screening ($\lambda_0=\infty$). 
The insets represent magnified views of the
transition region.}
\end{figure}

In the present model, as well as in  reality, the nature of 
fluctuations
along the  field (longitudinal direction) and 
perpendicular
to the field (transverse direction) could differ.
An extreme possibility here may be that the VG order occurs only in
some spatial component, say, in
the transverse component, keeping the other (longitudinal)
component  disordered \cite{Lopez}. Indeed, the possibility of
such ``two-dimensional'' 
or purely ``transverse'' vortex order has been discussed 
in the literature as a ``decoupling''
transition \cite{decoupling}. 
In order to probe  such exotic possibility,
we introduce  a transverse Binder  ratio in terms of
the layer-overlap $q'_k$ defined for the $k$-th $xy$-layer of 
the lattice by,
\begin{equation}
  q'_k = \frac{1}{L^2}\sum_{i\in k}
\psi_{i}^{(1)*}\psi_{i}^{(2)},
\end{equation}
\begin{equation}
  g_{trans}(L)=2-\frac{(1/L)\sum_k [\langle |q'_k|^4\rangle]}
    {(1/L)\sum_k [\langle  |q'_k|^2\rangle]^2}.
\end{equation}
When the VG order occurs in each layer, 
$g_{trans}(L\rightarrow \infty )$ should be nonzero. 
In particular, if the purely transverse VG order is to occur
as a consequence of the layer-decoupling, 
$g_{trans}(L\rightarrow \infty )$
should stay finite while $g(L\rightarrow \infty )$ vanishes.

The calculated $g_{trans}(L)$  
is shown in Fig.1(b).
As can be seen from Fig.1(b), $g_{trans}(L)$
exhibits behavior quite similar to $g(L)$,
revealing a
merging or a weak crossing at $T/J=0.67\pm 0.02$.
This indicates that the present model exhibits only a single bulk
VG transition where both the transverse and longitudinal components
order simulataneously.

\begin{figure}[h]
\label{fig:g-screening}
\epsfxsize=\columnwidth\epsfbox{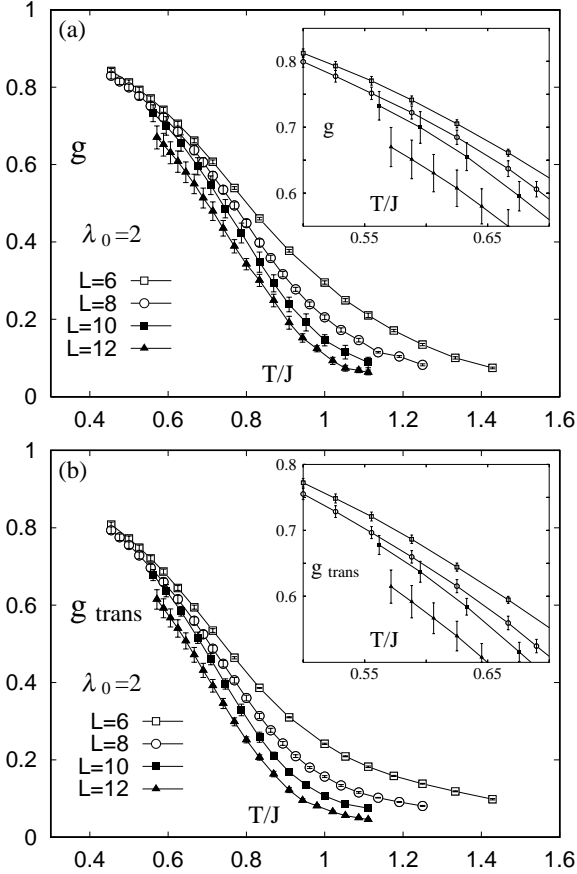}
\caption{
Temperature and size dependence of (a) the bulk Binder ratio
and  of (b) the transverse Binder ratio, of the  model with
screening ($\lambda_0=2$). The insets represent magnified views.
}
\end{figure}

In the case of  finite screening $\lambda_0=2$,
as shown in Figs.2(a) and 2(b), 
$g(L)$ and $g_{trans}(L)$
constantly decrease with increasing $L$ at all temperatures studied,
suggesting that a finite-temperature 
VG transition is absent in the presence of screening. 
Similar behavior is observed also
for the case of stronger screening, $\lambda_0=1$,
and it can be concluded that the screening effect destabilizes
the VG order at finite temperature. 
Thus, concerning the presence or absence of the VG order,
the present model yields the same answer as the 
spatially isotropic gauge-glass model.

Next, we turn to the critical properties of the model. 
For a finite-temperature VG transition for
$\lambda_0=\infty$, 
we estimate the correlation-legth exponent $\nu = 2.2\pm 0.4$
via the finite-size scaling analysis of $g(L)$
(see Fig.3). Then, from
the order parameter $q^{(2)}=[<q^2>]$, 
the critical-point-decay
exponent is  determined to be $\eta =-0.5\pm 0.2$. Likewise, 
from the decay of the autocorrelation function of 
the superconducting
order parameter at $T=T_g$ \cite{comment}, the dynamical exponent $z$ 
is estimated
to be $z= 3.3\pm 0.5$. 

\begin{figure}[h]
\label{fig:g-fss}
\epsfxsize=\columnwidth\epsfbox{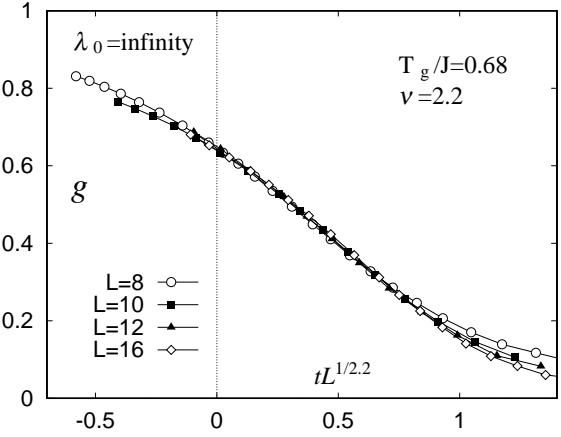}
\caption{
Finite-size scaling plot of the  bulk Binder ratio
of the  model without
screening ($\lambda_0=\infty$).
}
\end{figure}

In order to examine the possibility of
anisotropic scaling where
the longitudinal
and transverse correlation lengths have  different 
$\nu $ values,
we perform the scaling analysis  of $g_{trans}$ and
$q^{(2)}_{trans}$ as compared with $g$ and $q^{(2)}$ \cite{comment1}: 
If the scaling is really anisotropic, there is
no reason to expect that the
exponents determined from these transverse quantities coincide
with those determined from the bulk quantities.
We get $\nu \simeq 1.8$ and $\eta \simeq -0.5$ 
from $g_{trans}$ and $q^{(2)}_{trans}$, 
which agree  within the errors
with $\nu $ and $\eta $ determined above from $g$ and $q^{(2)}$.
We found no evidence of anisotropic scaling, although the occurrence of
small anisotropy cannot be ruled out from the present data.

Concerning the $T_g=0$ transition of $\lambda_0=2$,
we get  $\nu \simeq 4.0$ and 
$\nu \simeq 3.5$ from  $g$ and $q^{(2)}$, 
and $\nu \simeq 4.0$ and $\nu \simeq 3.0$ 
from $g_{trans}$ and $q^{(2)}_{trans}$, 
which again agree within the errors. (Note that $\eta $ should be equal to
-1 for a $T=0$ transition with nondegenerate ground state
as expected in the present model.)
Hence, the scaling also appears to be isotropic 
with $\nu =3.5\pm 1.0$.

The obtained exponents are summarized in  Table 1 and are compared
with the values  reported for the spatially isotropic
gauge-glass model with and without
screening. Apparently, there exists a significant deviation between
the two results.
In particular,
$\nu $ of the present model appears to be significantly 
larger than 
$\nu $ of the gauge-glass model \cite{comment3}, 
which might suggest
that the present model
lies in a universality class different from that of the gauge-glass model.

Finally, we wish to discuss the experimental implication of  our
results.  The present study suggests, 
in accord

\noindent
TABLE 1 
Critical exponents of the present model compared with the
values reported for the gauge-glass model for both cases 
of $\lambda_0=\infty$
(no screening) and $\lambda_0<\infty$ (finite screening).
\par\medskip

\begin{tabular}{l|l|c c c} \hline\hline
 & model & $\nu$ & $\eta$ & $z$ 
\\ \hline
$\lambda_0=\infty$ & present & $2.2(4)$ & $-0.5(2)$ & 
$3.3(5)$ 
\\ \cline{2-5}
 ($T_g>0$) & gauge glass & 
 \begin{tabular}{c} 
 $1.3(4)$  \cite{Reger} \\
 $1.3(3)$  \cite{Young} 
 \end{tabular} &
 & $4.7(7)$  \cite{Reger} 
\\ \hline
$\lambda_0<\infty$ & present& $3.5(10)$ & $-1$ & 
\\ \cline{2-5}
 ($T_g=0$) & gauge glass & 
 \begin{tabular}{c} 
 $1.05(10)$  \cite{Young}\\
 $1.05(3)$  \cite{Rieger}   
\end{tabular} &
 & 
\\ \hline\hline
\end{tabular}

\bigskip\bigskip

\noindent
with the previous studies based on the gauge-glass model,
that in random type-II superconductors with
point disorder there should
be no VG phase at finite temperature
in the strict sense. As  discussed,
experimental data on films
and twinned crystals supporting 
the occurrence of a thermodynamic transition into the truly
superconducting glassy state
might well reflect the properties associated with   
extended detects. In this connection, the properties of 
a sample  exclusively containing point defects
is of great interest [4-6]. 
Recently, Petrean {\it et al\/} measured the 
transport properties of such sample, 
untwinned, proton irradiated YBCO \cite{Petrean}.
These authors observed an Ohmic behavior at all temperature studied, 
but found that the linear resistivity appeared to vanish 
at a finite $T_g^*$ as 
$\rho_L \sim (T-T_g^*)^s$ with a {\it universal\/} exponent
$s\simeq 5.3\pm 0.7$. Since the non-Ohmic
region could not directly be reached in these measurements, it is
not clear at the present stage whether the non-Ohmic behavior
really sets in below the apparent $T_g^*$ deduced from the high-temperature
Ohmic regime. 
An another possibility, which is entirely consistent with the present 
result,  may be that the
power-law decrease of
$\rho _L$ eventually breaks down at some temperature 
close to 
$T_g^*$, yielding a small but nonzero $\rho _L$ even at $T<T_g^*$.
If this is the case, the universal
critical behavior of $\rho _L$ observed above
$T_g^*$ should be governed by the
$\lambda_0=\infty $ VG fixed point [1,20] which, however, should eventually be 
unstable against the screening effect.
Indeed, our present estimate of
$s=(z-1)\nu \simeq 5.1$ for the $\lambda_0=\infty $ transition
is close to the experimental value of Ref.\cite{Petrean}. It
might be interesting to experimentally determine the 
exponents $\nu $, $z$ and $\eta $ separately, as well as
to go further down to lower temperatures in order to examine
whether the non-Ohmic behavior really sets in below $T_g^*$.

In summary, we have introduced a VG model which possesses 
a uniform field and
cures the limitations of the  gauge-glass model.
Eextensive  simulations show
that, 
while the stable vortex-glass phase occurs 
in the absence of screening, it is eventually destroyed 
by the screening effect.
Critical exponents associated the VG transitions
appear to differ from those reported for
the gauge-glass model, suggesting that real VG transitions
may lie in a different universality class.

The numerical calculation has been performed on the FACOM VPP500
at the supercomputer center, ISSP,
University of Tokyo. 
The author is thankful to R. Ikeda
and  S. Okuma
for useful discussion.

\end{multicols}


\begin{thebibliography}{99}
\bibitem{comment}
In calculating the autocorrelation, we employed the standard
Metropolis dynamics without the temperature-exchange process.
\bibitem{comment3}
Note that estimates of Refs.\cite{Young,Rieger} for the gauge-glass model with 
screening were obtained in the strong screening limit $\lambda_0\rightarrow
0$, on the expectation that the exponents in the  
$\lambda_0\rightarrow 0$ limit should be the same as those for
$\lambda_0 >0$.
If the limit $\lambda_0\rightarrow 0$ were special contrary to 
this expectation,
it could be the cause of the observed discrepancy, at least partially.
\end{thebibliography}
\end{document}